\documentclass[twocolumn,numberedappendix,tighten]{emulateapj}
\usepackage{amssymb,amsmath,amsthm}
\usepackage{longtable}
\usepackage{amssymb}
\usepackage{amsmath, xspace}
\usepackage{graphics,graphicx} 
\usepackage{rotating}
\usepackage{color}
\usepackage{perpage} 

\usepackage[colorlinks=true,citecolor=blue ]{hyperref}
\begin{document}


\title{Searching for the 3.5 keV Line in the Deep Fields with Chandra: the 10 Ms observations}


\author{
Nico~Cappelluti\altaffilmark{1,2,3},
Esra~Bulbul\altaffilmark{4}, Adam~Foster\altaffilmark{5}, Priyamvada Natarajan\altaffilmark{1,2,3}, Megan C. Urry\altaffilmark{1,2,3}, Mark~W.~Bautz\altaffilmark{4}, Francesca~Civano\altaffilmark{5}, Eric~Miller\altaffilmark{4}, and Randall~K.~Smith\altaffilmark{5}
}
\altaffiltext{1}{Yale Center for Astronomy and Astrophysics, P.O. Box 208121, New Haven, CT 06520, USA.\\}

\altaffiltext{2}{Department of Physics, Yale University, P.O. Box 208121, New Haven, CT 06520, USA.\\ }

\altaffiltext{3}{Physics Department, University of Miami, Coral Gables, FL 33124  \\}

\altaffiltext{4}{Kavli Institute for Astrophysics \& Space Research, Massachusetts Institute of Technology, 77 Massachusetts Ave, Cambridge, MA 
02139, USA.\\}

\altaffiltext{5}{Harvard-Smithsonian Center for Astrophysics, 60 Garden Street, Cambridge, MA 02138, USA}

\begin{abstract}
In this paper we report  a systematic search for an emission line around 3.5 keV in the spectrum of the Cosmic X-ray Background using a total of $\sim$10 Ms Chandra observations towards the COSMOS Legacy and  CDFS survey fields. We find a marginal evidence of a feature at an energy of $\sim$3.51 keV with a significance of 2.5-3 $\sigma$, depending on the choice of the statistical treatment.
The line intensity is best fit at $8.8\ \pm\ {2.9}\times10^{-7}$ ph cm$^{-2}$s$^{-1}$ when using a simple $\Delta\chi^2$ or $10.2\ ^{+0.2}_{-0.4} \times10^{-7}$ ph cm$^{-2}$s$^{-1}$  when MCMC is used. Based on our knowledge of  $Chandra$, and the reported detection of the line by other instruments, an instrumental origin for the line remains unlikely. We cannot though rule out a statistical fluctuation and in that case our results provide a  3$\sigma$ upper limit at 1.85$\times$10$^{-6}$ ph cm$^{-2}$s$^{-1}$. We discuss the interpretation of this observed line in terms of the iron line background; S {\sc XVI} charge exchange as well as potentially  from sterile neutrino decay. We note that our detection is consistent with previous measurements of this line toward the Galactic center, and can be modeled as the result of  sterile neutrino decay from the Milky Way for the dark matter distribution  modeled as an NFW profile. For this case, we estimate a mass m$_{\nu}\sim$7.01  keV and a mixing angle sin$^2$(2$\theta$)= 0.83--2.75 $\times 10^{-10}$. These derived values are in agreement with independent estimates from galaxy clusters; the Galactic center  and M31.  

\end{abstract}

\section{Introduction}
\label{sec:intro}
Astrophysical and cosmological observations of gravitational interactions of visible baryonic matter provide overwhelming evidence for the existence of an additional dominant, component of non-luminous matter, referred to as dark matter \citep[see e.g.][]{rubin}. Extensive direct and indirect searches for this ubiquitous matter have so far failed to detect it, and, its nature remains unknown. The majority of this unseen component is inferred to be cold and collisionless, however, a warmer component can also be accommodated to account at least partially to the overall mass budget of dark matter. X-ray observations of dark matter-dominated objects, such as galaxies and clusters of galaxies, provide a unique laboratory for searching for the decay or annihilation of a viable warm dark matter candidate, namely sterile neutrinos \citep{dodelson94, dd02, abazajian01, boyarsky2006}. 

An unidentified emission line near 3.5~keV was recently detected in stacked observations of galaxy clusters and in the  Andromeda galaxy \citep[][Bul14a and Bo14 hereafter]{b14, bo14}. The interpretation of this signal as arising from decaying dark matter, has drawn considerable attention from astrophysics and particle physics communities. The line is also detected in the {\it Suzaku} and {\it NuSTAR}  observations of the core of the Perseus and the Bullet clusters \citep{wik,urban15,franse16} and in the Galactic center \citep{bo15}. An emission line at a consistent energy is also detected in  {\it XMM-Newton} observations of the Galactic center and in other individual clusters \citep{iakubovskyi15}. Recently, a 11$\sigma$-detection of the line was reported in summed {\it NuSTAR} observations of the COSMOS and Extended Chandra Deep Field South (ECDFS)  survey fields, where a dark matter signal from the Milky-Way halo may be expected \citep{nero}. As noted, another interesting  dark matter candidate that might also produce a 3.5~keV X-ray line is self interacting  dark matter from relatively low mass axion-like particles \citep[e.g., ][]{conlon14}.

Although the line was detected by several X-ray satellites, including {\it XMM-Newton, Chandra, Suzaku}, and {\it NuSTAR} in a variety of dark matter-dominated objects, several other studies  report non-detections of the line, e.g. in stacked {\it Suzaku} observations of clusters of galaxies \citep{bulbul16}, the dwarf galaxy Draco \citep{ruchayskiy16}, and {\it Hitomi} observations of the Perseus cluster \citep{hitomi16}. 
However, the upper limits derived from the stacked galaxies are  in tension with the original detection at the $5\sigma$ level \citep{anderson15}. 

Despite these intensive and persistent efforts, the origin of the 3.5~keV line remains unclear. Potential astrophysical interpretations were discussed extensively by Bul14. A more recent update is provided by \citet{franse16}, who consider an additional model that comprises a charge exchange between bare Sulfur ions and neutral gas (e.g., Bul14a, Gu et al. 2016, Shah et al. 2016). The radial distribution of the flux of the line can provide an independent test of its origin; however, the observed line flux from the Perseus core is consistent with a dark matter origin \citep{franse16}. However, the intensity of the signal in the cluster core appears to be anomalously high for the decaying dark matter model (Bul14a, Franse et al.\ 2016). In their recent paper, the {\it Hitomi} collaboration measure  the K {\sc XVIII} abundance for the first time as 0.6Z$_{\odot}$, well within the allowed limits in Bul14 in the core of the Perseus cluster \citep{hitomi16}.  The other possible astrophysical line which was suggested as a contaminant is Ar {\sc XVII} DR from lab studies of Electron Beam Ion Trapping measurements \citep{bulbul17}. These  results have eliminated K {\sc XVIII}  and Ar {\sc XVII} DR lines as the possible origin for the 3.5 keV line.  The $Hitomi$ collaboration  reports tension between the flux in the Perseus cluster observed by {\it XMM-Newton} and {\it Hitomi} at the 3$\sigma$ level. The authors attribute this discrepancy to subtle instrumental features in earlier observations of {\it Hitomi}. 

Here, we report the detection of the  line at $\sim$3.5~keV in the summed data from deep {\it Chandra} blank fields, the {\it Chandra} Deep Field South (CDFS) and COSMOS for a total exposure of 9.17 ~Ms. We critically discuss instrumental effects together with four  plausible explanations for the origin of the 3.5~keV line  --  charge exchange; the iron line background; a statistical fluctuation and dark matter decay. All errors quoted throughout the paper correspond to 68\% single-parameter confidence intervals. Throughout our analysis we use a standard $\Lambda$CDM cosmology adopting the following values for the relevant parameters: $H_{0}$ =71 km s$^{-1}$ Mpc$^{-1}$, $\Omega_{M}= 0.27$, and $\Omega_{\Lambda}= 0.73$.

\section{Data sets} 
The {\em Chandra}-COSMOS Legacy Survey \citep[hereafter, CCLS;][]{scov,elvis,civ16} and
the {\em Chandra}-Deep Field South \citep[hereafter CDFS;][]{giacc01,luo08,x11,luo16}
 have been observed for  $\sim$4.6 M and $\sim$7 Ms  respectively,  with the ACIS-I CCD  instrument onboard  {\em Chandra} with  117, and 111  pointings, respectively. The CCLS field is a relatively shallow mosaic of $\sim$2 deg$^2$ and an  average exposure 
 of $\sim$160 ks/pix while, the CDFS field is a deep pencil beam survey of $\sim$0.1 deg$^2$ observed for 7 Ms/pix. However, since the signal  is 
 very faint, for spectral analysis we have only used the pointings observed in the VFAINT telemetry mode with a focal 
 plane temperature of 153.5 K, in order to minimize the instrumental background. Since the CDFS was
 partly observed in the early phase of the mission when the VFAINT mode wasn't available and observations were partly taken 
 at higher temperature, the total exposure time before treatment is $\sim$6 Ms. 
     
 \section{Data Analysis}

 Raw event files were calibrated using the CIAO tool $chandra\_repro$ and the Calibration Data Base (CALDB) version 4.8. For every pointing, 
 time intervals with high background were cleaned using the CIAO tool $deflare$ using the $lc\_clean$ technique as described 
 by \citet{hm06}. The de-flaring was performed in the [2.3-7] keV, [9.5-12] keV and [0.3-3] keV energy band in sequence, in 
order to detect flares with anomalous hardness ratios \citep{hm06}. Although not critical for this work, the astrometry was aligned 
using reference optical catalogs. 

The X-ray signal is a blend of detected and unresolved AGN, galaxies and clusters whose summed emission is often referred to as the 
"Cosmic X-ray Background" (CXB).
There is also a particle-induced background and a (relatively small ) background from other sources within the instrument.  
 Hereafter we will use the acronym CXB for the signal produced by all astrophysical sources that is focused by the 
optics and we adopt the acronym PIB for the "Particle and Instrumental Background " which is produced by all other (non-astrophysical) sources.

For sake of clarity, in this paper, the putative 3.5 keV signal arising either from dark matter decay or Sulfur charge exchange, will be considered as a separate 
component on top of the CXB and PIB signal. Therefore, we start the analysis by carefully accounting for known X-ray sources, that constitute the 
PIB and the CXB. 
\begin{figure*}
\begin{center}
\includegraphics[width=7in]{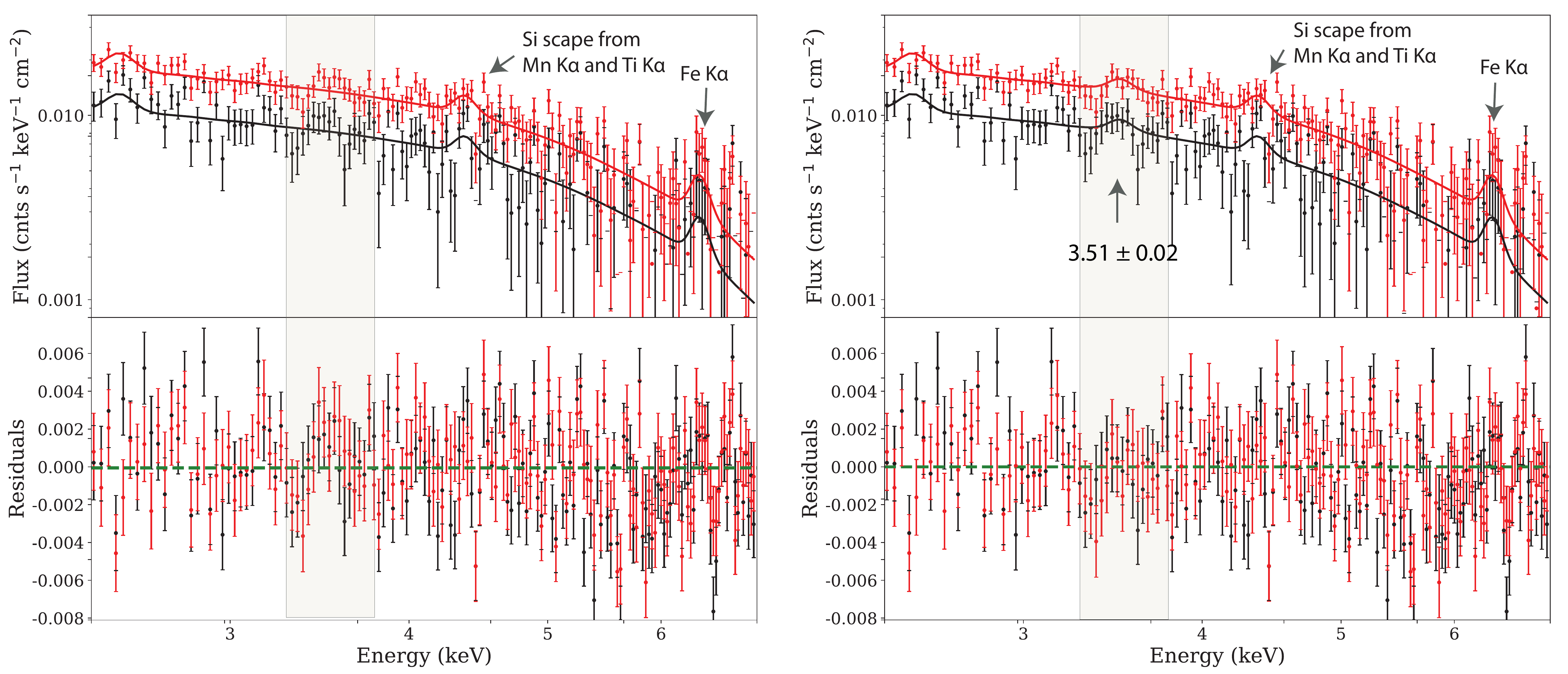} 
\caption{$\bf Left\ Panel:$  The CXB spectra in the  CDFS ($in\ black$) and CCLS ($in \ red$) together with best-fit models (solid lines) and the residuals without the 3.5 keV line  Gaussian model component.
$\bf Right \ Panel:$ the same by adding a Gaussian model at $\sim$3.5~keV. The known instrumental lines Si escape peak from Mn  K$\alpha$, Ti K$\alpha$ at 4.4 keV and Fe K$\alpha$ at 6.4 keV are marked in both panels. 
\label{fig:cxb}.
}
\end{center}
\end{figure*}

\begin{figure*}
\begin{center}
\includegraphics[width=.9\textwidth]{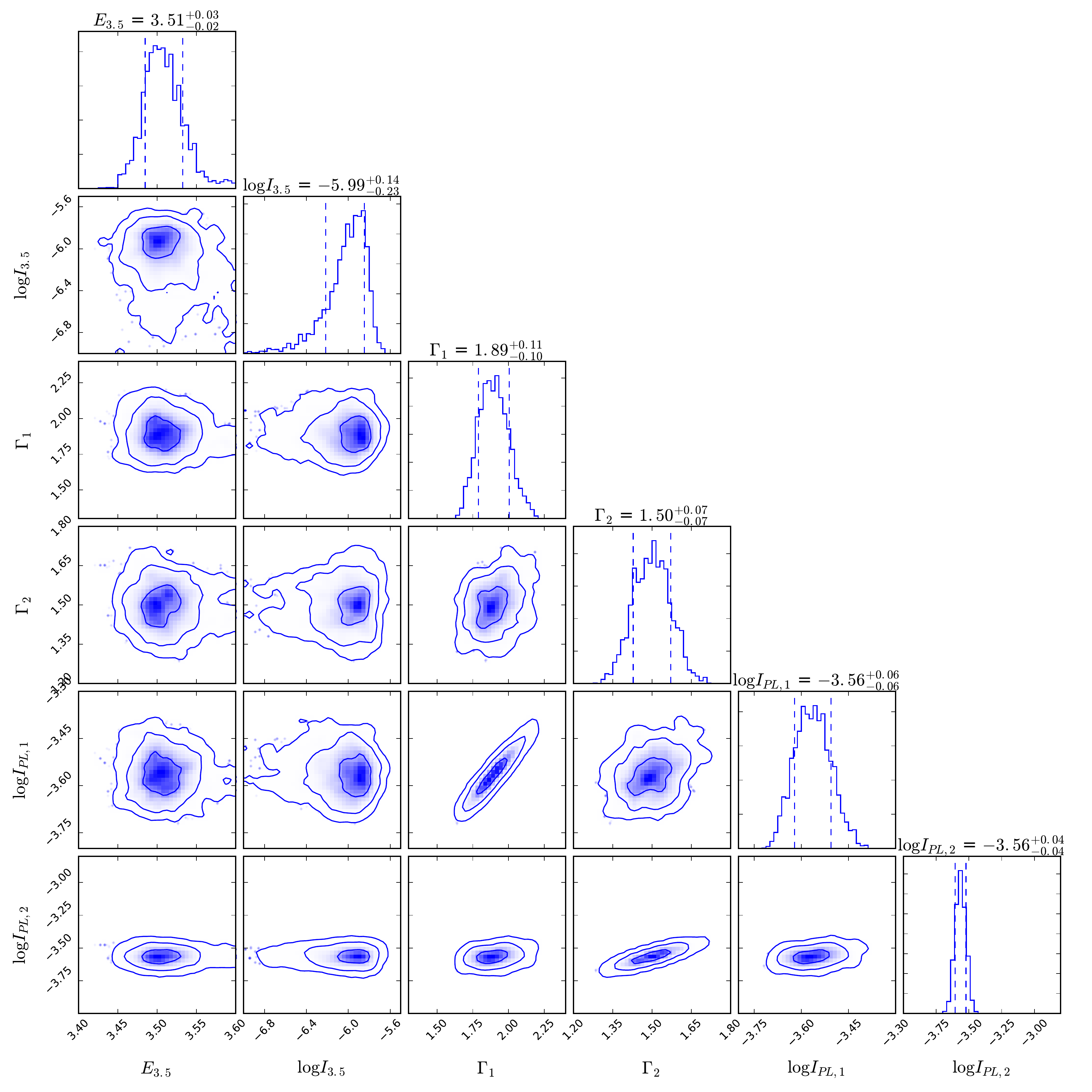} 
\caption{In blue color scale: the fit parameters confidence contours for the background   subtracted full detector case obtained with the MCMC analysis for the relevant parameters. 
The contours levels are 1,2 and $\sigma$, respectively. For every parameter we plot the marginal probability distribution  histogram on top of every column where we show the 1$\sigma$ intervals with dashed lines. We also report the best-fit values and the 1$\sigma$ confidence level. \label{fig:mcmc1}. From top to bottom the parameters are: Energy (E$_{3.5}$) and logarithm of the intensity of the 3.5 keV line ($log(I_{3.5}$)), 
the CXB spectral index in the CDFS $\Gamma_1$, the spectral index in the CCLS $\Gamma_2$, the  logarithm of the normalization of the continuum in the CDFS  ($log(I_{PL,1}$))and in the CCLS ($log(I_{PL,2}$)). \label{fig:nosrccont}}
 \end{center}
\end{figure*}

\begin{figure}
\begin{center}
\includegraphics[width=.4\textwidth]{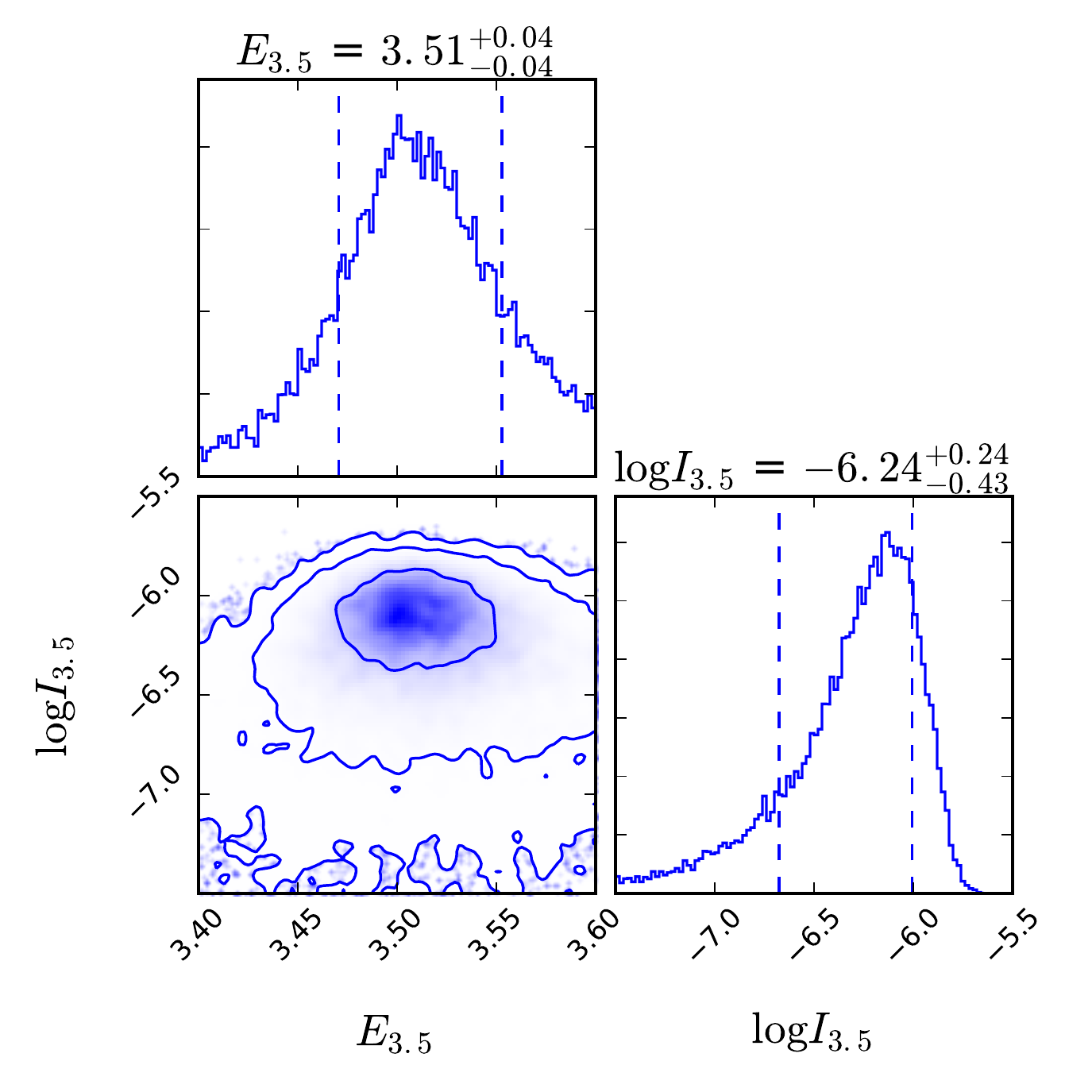} 
\caption{In blue color scale: the fit parameters confidence contours for the background   subtracted source masked case obtained with the MCMC analysis for the  3.5 keV like parameters,  intensity of  line ($log(I_{3.5}$) and Energy (E$_{3.5}$). The contours levels are 1,2 and $\sigma$, respectively.
 \label{fig:mcmc1}}
 \end{center}
\end{figure}

\subsection{Extraction of Summed X-ray Spectrum}
The detected intensity of the CXB is not the same across the surveys presented here, primarily due to cosmic variance,  so we derive an indipendent spectrum for each survey field. For each pointing, we extracted the spectrum of all the photons detected 
in the ACIS-I field of view (FOV) with the CIAO tool $specextract$.
 For each spectrum, we then computed the field-averaged Redistribution Matrix Functions (RMFs) and Ancillary Response Functions (ARF) using the CIAO-tool $specextract$. Spectra were  
 co-added and response matrices averaged after weighting by the exposure time.  We  produced  a cumulative CXB+PIB  spectrum for each of the  datasets. Because we are looking for diffuse emission,
 the only background component in our observations is the PIB. The {\em Chandra} X-ray observatory periodically obtains "dark frames", i.e. exposures  with ACIS in the stowed mode.
When the High Resolution Camera is on the focal plane, ACIS is stowed and unexposed to any focused source but it still records the PIB component. In such a position the ACIS detectors see neither the sky nor the calibration sources.
In particular, \citet{hm06} demonstrated that the [2-7] keV to [9.5-12] keV hardness ratio is constant (within 2\%) in time regardless of the amplitude of the particle background. Therefore, we employed ACIS-I observations in the stowed mode to evaluate the background.  In particular, we merged the $stowed$ mode event files, applied the VFAINT filtering and reprojected to the same astrometric frame as the observations, We then extracted the spectrum in the same source-masked regions and renormalized it by the ratio of count rate in energy bins C$_{[9.5-12],obs}$/C$_{[9.5-12],stow}$ where C$_{[9.5-12],obs}$ and C$_{[9.5-12],stow}$.
In a recent paper, \citet{barta} performed a detailed and sophisticated analysis of the  same stowed ACIS-I event files employed here and reported, to within 2\%, the relative stability of the background 
in observations of later epochs than those used by \citep{hm06}. In this paper, we  are looking methodically for astrophysical emission lines in the energy range [2.4-7] keV. In this energy band, the PIB is affected 
by a systematic uncertainty of the order of  2\% which is added in quadrature to the PIB spectral data error bars throughout our analysis. In Table \ref{tab:cts}, we summarize the number of net counts used for the spectral modeling and the resulting vignetting-weighted final exposures for our datasets. However, we note that the observations in the stowed mode are much shorter than the those employed here (a total of 1 Ms in the archive vs 9.16 Ms). This of course, significantly limits our sensitivity, since the PIB spectrum has larger errors than those in the data and therefore might potentially artificially smooth out any features in the data. 
\begin{table}
\begin{center}
\footnotesize
\caption{[2.4-7] keV net counts and exposures\label{tab:cts}}

\renewcommand{\arraystretch}{1.5}
\begin{tabular}{lccc}
\hline\hline
	& SIGNAL & BACKGROUND &EXPOSURE\\
	&       (counts)          & (counts)        & Ms \\
	\hline\hline

CDFS &    115373     &          1989189 &  5.57   \\
CCLS  & 131826       &          1220611 & 3.59         \\
Total &    247199 & 3209800& 9.16\\ 
\hline\hline
\\
\end{tabular}
\end{center}
\end{table}

\subsection{About the spectrum of PIB \label{Sect:PIB}}
\label{sec:pib}

Part of the signal included in the total X-ray spectrum is due to the PIB. In order to find faint sources and/or to analyze faint, diffuse emission-lines, careful treatment of these backgrounds is essential. We start by examining data from ACIS-I in stowed mode, i.e., when no cosmic photons are collected. This provides a robust representation of the particle background plus internal instrumental background. Although an universal model of the PIB is not provided by the {\em Chandra} team, here we can model the PIB using a broken power-law, with the slopes ($\Gamma_{PIB,1}$, $\Gamma_{PIB,2}$); the break energy  (E$_{break}$)  and the normalization ($norm$) as free parameters. On top of this, we add a Gaussian model at $E\sim2.5$~keV  and  $E\sim$5.9~keV, with energies and intensities (I$_{1,2}$) that are free to vary. The  line at 5.9 keV is a known Mn K$_{\alpha}$ instrumental feature. This feature is scattered light from the radioactive $^{55}$Fe in the external calibration source. This source has a half-life of $\sim$2.7 years, so its intensity has dropped dramatically over the course of the Chandra mission. So its not surprising that it (and its K-escape and Ti line) is not fully subtracted from the CXB spectrum. 
The line at 2.51 keV  is instrumental and an artifact: \citet{barta} pointed out that in the [2-3] keV energy band, due the position dependent charge transfer inefficiency (CTI) correction the strong broad emission line at $\sim$2.1 keV ($mother~line$)  produces a system of spurious $daughter~lines$ at energies of up to $\sim$2.6 keV along with spurious broadening. A similar effect is observed above 7.3 keV as well. CTI correction is necessary because radiation has damaged the ACIS-I resulting in 
loss in the charge transfer inefficiency. This damage however did not affect areas of the CCD not exposed to the X-rays such as the frame store area. To cope with the CTI, a correction is applied  $a~posteriori$ by the data analysis pipeline. This correction is applied to all the data including those collected by areas not damaged by radiation. The result is that for the strongest instrumental emission lines, the recorded energy is artificially shifted up to 800 eV higher energy (depending on the position on the detector). Detailed modeling of PIB is beyond the scope of this paper an we refer the readers to the {\em Chandra}~Calibration~Database and to specific papers \citep[see e.g.][]{barta}.

%
%


\section{Results}
As noted in Table \ref{tab:cts} the spectra analyzed in this paper are background dominated, and this 
might raise  concern when looking for faint emission lines.
In this section, we present two different approaches to present the results based on two indipendent  methods to handle the background. In the first we subtract the properly normalized PIB spectra from the data and fit  and in the second, we fit the CXB+PIB at the same time with models for each component.  If we detected the 3.5 keV line if its diffuse  coming from  the entire of view, we performed the fit using data accumulated over the whole detector and after masking the detected  sources.

\subsection{Fitting the Background Subtracted Full Spectra Including Point Sources}

XSPEC v12.9.0 was used to perform the spectral fits with $\chi^2$ as an estimator of the goodness-of-fit. The spectral counts in each energy bin are sufficient to allow the use of the Gaussian statistics in this analysis \citep{protassov02}. To increase the sensitivity to weak emission lines, we simultaneously fit the CXB spectra from the CCLS and CDFS.  We restrict the energy range to  2.4--7 keV in order to avoid the bright $Au$ feature at 2 keV, while having sufficient leverage on the power-law component.  
The Galactic column densities are fixed to 2.5$\times10^{20}$ cm$^{-2}$ for the fits of the CCLS field and 8.8$\times10^{19}$ cm$^{-2}$ for the CDFS field \citep{dickey1990}. The power-law indices and  the normalizations are left free in our fits to account for the different CXB flux in the two fields \citep{hm06}. We first fit the spectra with a single absorbed ($wabs$ model in XSPEC) power-law model which gives an overall good fit with $\chi^{2}$ of 563.43 for 308 degrees-of-freedom (dof). 

The best-fit power-law normalizations are found to be: $\sim$2.78 $\times 10^{-4}$ ph keV$^{-1}$ cm$^{-2}$ s$^{-1}$ in CDFS and 2.80$\times 10^{-4}$ ph keV$^{-1}$ cm$^{-2}$ s$^{-1}$ in CCLS. The power-law indices are $\Gamma_1$=1.82$\pm{0.10}$ and $\Gamma_2$=1.48$\pm{0.06}$ for the two fields, respectively (hereafter the subscripts 1,2 will refer to CDFS and CCLS respectively). 
The fluxes and spectral indices measured here  are in agreement with \citet{hm06}, \citet{more}, \citet{barta} and \citet{cap17}.

A few spectral features are immediately visible around 2.51 keV, 3.15 keV, 3.5 keV,  4.4 keV, and 6.4 keV.   
The 2.51 keV line is a strong $Au$-M complex line. We tried to fit the feature at 3.15 keV and didn't find a significant line but only found 
a 3$\sigma$ upper-limit of $\sim$1.5$\times 10^{-6}$ ph keV$^{-1}$ cm$^{-2}$ s$^{-1}$, this means that the feature is just a statistical fluctuation in a few channels.
The emission line at 4.37 keV is consistent with a residual from a blend of known instrumental emission lines from Silicon escape (i.e. lines formed by  electron clouds left when a  photon carrying away energy leaves silicon substrate)\footnote{\url http://cxc.harvard.edu/cal/Acis/Cal\_prods/matrix/notes/Fl-esc.html} from Mn K$_{\alpha_{1,2}}$ and Ti K$_{\alpha_{1,2}}$ \footnote{\url  http://www2.astro.psu.edu/xray/docs/cal\_report/node155.html} given that the energy resolution is $>$200~eV at these energies. 
These two weak emission lines are hard to detect in the PIB due to limited statistics but they become clearly visible in the deep blank-sky observations used here or can be produced by a minimal leaking of the on board calibration source. 
The line at 6.4 keV is consistent with  Fe K$_\alpha$ and for this line we cannot discriminate between  an instrumental or a Galactic origin.
Adding the Gaussian components for the instrumental lines at  2.51 keV, 3.15 keV, 4.4 keV, and 6.4 keV with variable energies and normalizations improves the $\chi^{2}$  value by a significant amount with $\chi^{2}$ of 527.01 for 298 dof. 

{
\begin{table}[!b]
\begin{center}
\footnotesize
\caption{Best-fit emission line parameters from the Joint Fits of Deep Field CXB Spectra obtained
withm the MCMC method.  }
\renewcommand{\arraystretch}{1.5}
\begin{tabular}{lcc}
\hline\hline
Energy  	& Flux \\
keV             & 10$^{-6}$ ph  cm$^{-2}$ s$^{-1}$\\
\hline		
\hline		
   2.51 $\pm$ 0.01& 52.80 $\pm$ 19.64	 \\
   3.51 $\pm$ 0.02 & 1.02 $\pm$ 0.41 \\
   4.37 $\pm$ 0.03 & 1.12 $\pm$ 0.29	\\
   6.38 $\pm$ 0.04 & 1.98 $\pm$ 0.55	\\
\hline		
$\chi^2$ (dof) 			&   527.01  (298)	 

\\
\hline\hline
\\
\multicolumn{1}{l}{%
  \begin{minipage}{6.cm}%
  \end{minipage}%
  
}\end{tabular}
\label{tab:sfree}
\end{center}
\end{table}
}
 {
\begin{table}[!b]
\begin{center}
\footnotesize
\caption{Best-fit Model Continuum of the two  fields CXB Spectra.  }
\renewcommand{\arraystretch}{1.5}
\begin{tabular}{lcc}
\hline\hline
Parameter  	& Value & Unit\\
          & \\
\hline		
\hline		
$\Gamma_1$&1.89 $_{-0.10}^{+0.10}$	& \\
$\Gamma_2$&1.50 $_{-0.07}^{+0.07}$	 &\\
log($I_{PL,1}$)& -3.56$\pm{0.01}$ &ph cm$^2$ s$^{-1}$\\ 
log($I_{PL,2}$)& -3.56$\pm{0.04}$& ph cm$^2$ s$^{-1}$\\ 
\hline		

\\
\hline\hline
\\
\multicolumn{3}{l}{%
  \begin{minipage}{6.cm}%
  \end{minipage}%
  }
\end{tabular}
\label{tab:cont}
\end{center}
\end{table}

We present the data and the best-fit model obtained with (right panel) and without (left panel) a Gaussian line added in the model at 3.5 keV line in Figure\ref{fig:cxb}. 
The best-fit energy of the Gaussian at 3.5 keV line becomes 3.51$_{-0.02}^{+0.02}$ keV with a flux of 8.83$\pm{2.9}\times 10^{-7}$ ph cm$^{2}$ s$^{-1}$. If this line is removed from the fit the change in $\chi^2$ value becomes 536.93 ($\Delta\chi^{2}$ of 10.23) for 2 dof, corresponding to a detection confidence level of 3.2$\sigma$.  From the $\chi^2$ contour we determine a 3$\sigma$ upper limit of 1.75 $\times10^{-6}$ ph cm$^{2}$ s$^{-1}$.
 This  would  correspond to  P$\sim$0.003 (i.e. probability that the line is not present). However, in cases like this  the model is not correctly specified (the best fit should have had $\chi^{2}\sim$298): when the model is misspecified, the traditional correspondence between $\Delta\chi^2$ and P breaks down \citep[see e.g.][] {span}.  To fully understand the actual level of P one would need to perform more detailed tests that, because of the statistics, we did not perform in this work. However, we tested if the addition  of the emission lines improved the quality of fit with the Bayes  and  Aikake Information Criterion \citep{bic} (BIC and AIC, respectively). The change in BIC value is $\sim$15 while the AIC suggest that the Power only fit  is $\sim$10$^6$ times less likely 
than the Power-law plus emission lines model. However, when the BIC is computed between the power-law model  and the power-law plus any single detected emission line the quality of the fits are marginally improved. This is indeed one of the limitations of BIC
that tends to discard more complicated models and is not sensitive to low signal-to-noise Ratio signals.
The AIC instead always favors the power-law plus emission lines. 
We also use the Markov-Chain Monte-Carlo
(MCMC) solver in XSPEC to determine the full probability
distribution of the free fit parameters including the instrumental lines. Using the Metropolis-
Hastings algorithm, we run 5 chains, each with
a length of 25,000 and discard the first 5000 steps in each run for the burn-in period. 
Integrating over all the parameter we obtain the posterior distribution for each variable parameter ({P\it(X)}). 
Figure \ref{fig:mcmc1} shows the derived P(X) for each
 parameter (excluding instrumental lines) and the confidence contours and the best-fit parameters. 
 The best-fit power-law continuum parameters are $\Gamma_1$=1.89$_{-0.11}^{+0.10}$ , $\Gamma_2$=1.50$_{-0.07}^{+0.07}$ and fluxes Log(I$_{PL,1}$)=-3.56$\pm{0.01}$  ph cm$^{2}$ s$^{-1}$ and  Log(I$_{PL,2}$)=-3.56$\pm{0.04}$  ph cm$^{2}$ s$^{-1}$ in agreement with \citet{hm06,cap17}.
The continuum paremeters best-fit are summarized in Table \ref{tab:cont}, note that in this case flux is accumulated on a 16.9\arcmin$\times$16.9\arcmin area.
 The best-fit energy and flux of the 3.5 keV line are  consistent with those obtained with the $\chi^2$ fit and are E=3.51$_{-0.02}^{+0.03}$ keV and I$_{3.5}= 10.2_{-0.4}^{+0.2}\times 10^{-7}$ ph cm$^{2}$ s$^{-1}$.  P(I$_{3.5}$) is very asymmetric with a tail toward low values  floored at 3$\sigma$ at  7.2$\times$10$^{-8}$ ph cm$^{2}$ s$^{-1}$ hence confirming the significance of the line detection at $\sim$3$\sigma$ confidence.  In Table \ref{tab:sfree} we report all the detected emission lines parameters.	 The 3$\sigma$
upper limit found with MCMC is 1.85$\times10^{-6}$ ph cm$^{2}$ s$^{-1}$.
However we point out that like in the $\chi^2$ fit case,  MCMC can only reflect statistical variations, and does not treat model misspecification.  This  problem will be approached in a forthcoming paper that will employ a larger sample.  

 \subsection{Fitting the Background subtracted,  Source Masked Spectra}

As a further test we fit the spectrum obtained after masking all the known point and extended sources in the field.  At the time of the analysis the latest public catalog of CDFS sources was produced with the 4~Ms exposure of \citet{x11}. We mosaic all the available observations and produce exposure maps as described by \cite{cap16a}. We then run a CIAO's source detection algorithm {\it wavdetect} in the [0.5-2] keV, [2-7] keV, and [0.5-7] keV energy bands. We set a threshold of 10$^{-5}$ (see CIAO detect manual) and the faintest detected sources have fluxes of the order 10$^{-17}$ erg cm$^{-2}$ s$^{-1}$. For each point and extended source, we create regions with spatial extent of 5$\sigma$ of the PSF around the centroid  (ranging from $\sim$1-1.5$\arcsec$  full-width-half-maximum at the center of the image to $>$5$\arcsec$ at the outskirts).  The three-band catalogs are merged and sources in each of the bands are removed  from the event files of each pointing\footnote{We note that there is a substantial agreement with the  \citet{luo16} CDFS 7 Ms catalog that became available after the submission of this paper}. 

CCLS has a completely different tiling of pointings. Therefore, source detection requires a more complicated procedure. For CCLS we employe the catalog published by \citet{civ16} and mask sources within $\sim$10$\arcsec$ around each detection. According to Figure 9 of  \citet{civ16}, this procedure will safely remove $>$90\% of the sources' flux in the energy bands investigated here. An emission line with a best-fit energy of  3.51 keV  is detected at  2.5$\sigma$ confidence level.  
Although less strongly than above, even in this configuration the BIC and the AIC still favor the power law plus emission lines and the confidence contours obtained from MCMC analysis are shown in Figure \ref{fig:nosrccont}. The  best-fit energy and flux parameters found in the MCMC analysis are E=3.51$_{-0.04}^{+0.04}$ keV  and I$_{3.5}$=5.8$_{-3.8}^{+4.6}\times$10$^{-7}$  ph cm$^{2}$ s$^{-1}$ respectively. The line energy is poorly constrained while the intensity has been found larger that zero $>$97\% of the times, therefore providing an evidence for the line at around 2.5$\sigma$.

The best-fit energy and flux found in source included and source excluded spectra are in agreement within 1$\sigma$ level. The detection of the 3.5 keV line in the source excluded fit is less stringent than the source included case. This is due to the fact that the source masking, especially in the CDFS, removes a larger fraction of the data ($>$50\%) and hence the statistics on the continuum is severely affected. Therefore, the power-law and the flux of the 3.51 keV line continuum is weakly constrained. Since we detect the line in both source included and source excluded spectra at consistent energy and flux, this points to the fact that the signal is not resulted in from the point sources in the field, rather, it is extended in origin.

 \subsection{Fit the Spectra with a Background Model}
  
 We have then fitted the data plus background using two  models at the same time for a) the PIB described in  Sect. \ref{Sect:PIB} without folding it  through the ARF plus b) the CXB model desired the the previous two subsections folded through all the response matrices.
Since the PIB for the two datasets differs only in amplitude, in the PIB models the slopes ($\Gamma_{PIB,1}$, $\Gamma_{PIB,2}$) and  the break energy  (E$_{break}$)  parameters were tied while,  the normalizations  ($norm$) were set as  independent parameters.  
  On top of it we added the instrumental emission lines mentioned in Sect. \ref{Sect:PIB}  CXB component approximated with a power law absorbed with Galactic N$_H$ plus we tested the presence of the 3.5 keV line. Overall the model consists of 36 parameters, 
  therefore given the number of data points here any BIC or AIC test are meaningless \citep{bic}.
  
  The fit results are reported  in Figure \ref{fig:bkg_mdl} together with confidence contours obtained with MCMC. The line is detected with a significance of $\sim$2.5$\sigma$ with a lower, but still consistent, energy with respect to the case of the background subtracted case (I$_{3.5}$=3.9$^{-2.1}_{+2.5}\times$10$^{-7}$  ph cm$^{2}$ s$^{-1}$ and E=3.49$_{-0.03}^{+0.04}$ keV). Also in this case the probability 
   distribution for I$_{3.5}$ is skewed toward low values.
However, in  this fit we find an inconsistency between the CXB power-law normalizations  I$_{PL,1}$ and I$_{PL,2}$ with those 
reported above. The reason is because by fitting in the [2.4-7] keV energy range the software does not  have a mean to disentangle 
between the PIB and CXB power-laws normalization: I$_{PIB,1}$, I$_{PIB,2}$, I$_{PL,1}$, I$_{PL,2}$, respectively. Indeed the $\chi^2$ fit doesn't find a satisfactory value of the CXB spectral indices  as most of the signal is spuriously attributed to the PIB. We decided to freeze the CXB spectral indices to $\Gamma_{1,2}$=1.4 \citep{cap17}. Most of continuum parameter are highly covariant. For this reasons and for the complexity of the model we decided to rely on the background subtracted scenarios that provide a more
stable and  model independent result. For the same reason we do not show the source masked, background modeled, scenario.

 \begin{figure*}
\begin{center}
\includegraphics[width=\textwidth]{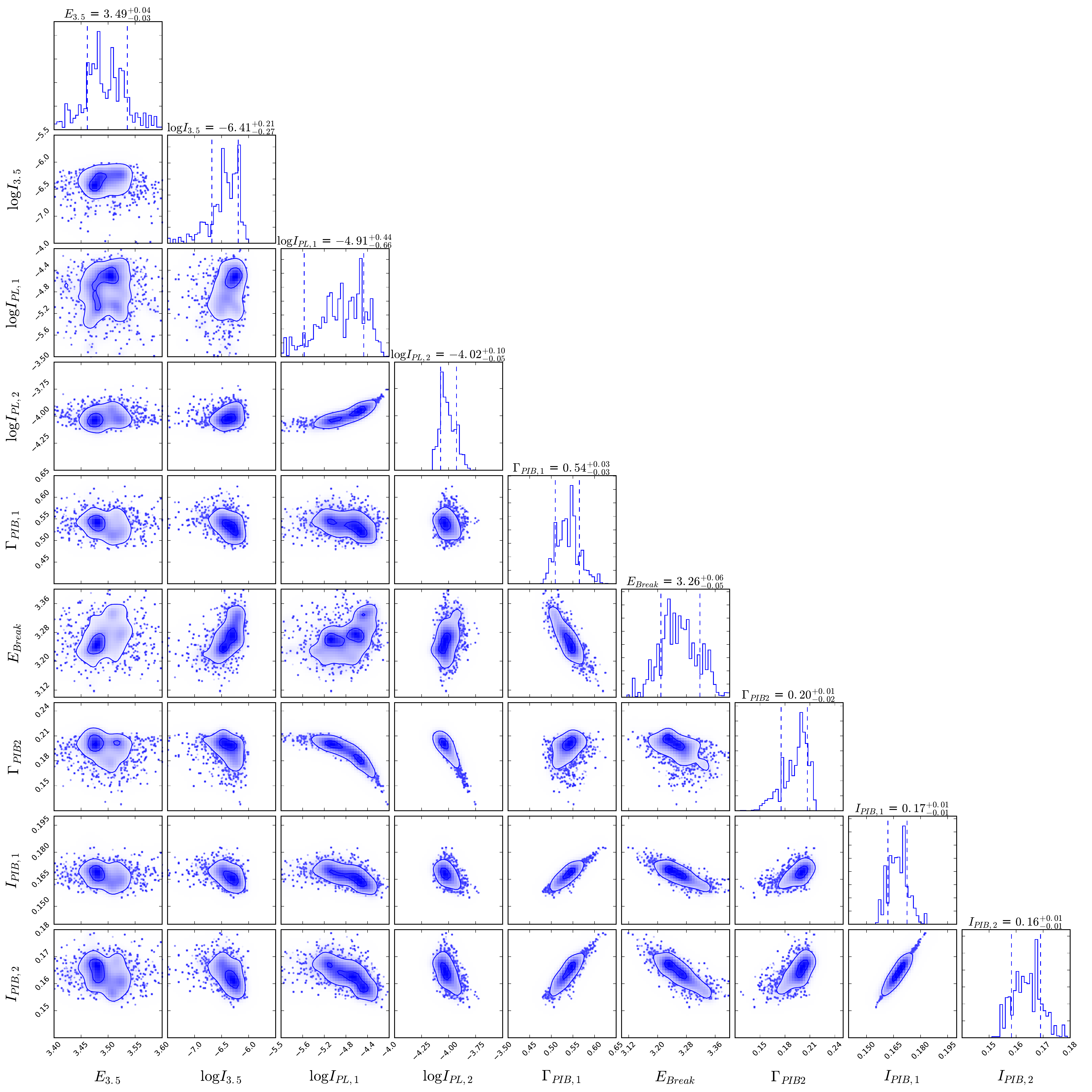} 
\caption{In blue color scale: the fit parameters confidence contours for the background   modeled full detector case obtained with the MCMC analysis for the relevant parameters. 
The contours levels are 1,2 and 3$\sigma$, respectively. For every parameter we plot the marginal probability distribution  histogram on top of every column where we show the 1$\sigma$ intervals with dashed lines. We also report the best-fit values and the 1$\sigma$ confidence level. From top to bottom the parameters are: Energy (E$_{3.5}$) and logarithm of the intensity of the 3.5 keV line ($log(I_{3.5}$)), the CXB logarithm of the normalization of the continuum in the CDFS  ($log(I_{PL,1}$)) and in the CCLS ($log(I_{PL,2}$)), the PIB spectral index in the CDFS ($\Gamma_{PIB,1}$) the  break energy (E$_{Break}$), the PIB spectral index in the CCLS ($\Gamma_{PIB,2}$), 
 \label{fig:bkg_mdl} the  break energy E$_{Break}$,  the PIB logarithm of the normalization of the continuum in the CDFS  ($log(I_{PIB,1}$)) and in the CCLS ($log(I_{PIB,2}$)).} . \label{fig:mcmc2}
 \end{center}
\end{figure*}


 %

\subsection{Safety tests}


%



%

Considering the marginal significance of the detection we asked ourselves if  the detected 3.5 keV line was a statistical fluctuation? As far as the 3.5 keV line is concerned,  this is not a blind search since the energy of
line under investigation is known a-priori.  This means that the {\it look elsewhere effect} in our measurement is not important or at least negligible.  However, given the low Signal-To-Noise ratio of the detected
signal, we   tested the hypothesis that the observed line might be a statistical fluctuation in the background. In order to test this, we obtained 1000 random realizations of the best-fit spectrum without the 
3.5 keV line via Monte-Carlo integration.  At the same time we also drew 1000 random realizations of the stowed background spectrum. With these datasets in hand we fitted every realization with the model including 
the 3.5 keV line and compute the cumulative distribution of the E$_{3.5}$ and I$_{3.5}$ fit results and found that, while the values of E$_{3.5}$ are uniformly distributed between 3 and 4 keV,  (3$\sigma$) the 3.5 keV line flux is always $<<1.0\times10^{-6}$ ph cm$^{2}$ s$^{-1}$, in agreement with our findings.  However since the background level is known with a $\sim$2\% precision, we cannot at the moment exclude that systematic effects could indeed produce the observed line but we point out that in the [3--4] keV band the overall spectrum is rather flat and the effective area is rather smooth. We also stress the fact that such a simulation is sensitive to statistical fluctuations only and not to systematics effects which, in this case, can be only estimated.

\section{Discussion}

We discuss our findings in the context of earlier claims of detection of the 3.5 keV line by several other groups. The 3.5 keV line has been previously 
detected in the \citet{} in the direction of the Perseus Cluster; in a stack of galaxy clusters, and more recently, toward the Galactic Center and  in M31 by Bo14. Interestingly, the energy of the line is consistent with that detected in Perseus redshifted from z=0.018.
However, the 
recent non-detection by $Hitomi$ \citep{hitomi16} rules out the highest flux detected by XMM-MOS in the the direction of Perseus. 
Recently, \citet{perez} made independent NuSTAR observations that are also relevant for testing the possibility of a 3.5-keV signal.    They found a significant line flux at 3.5 keV.  However, they also detected the line in observations where the Galactic Center direction is blocked by Earth.  As the nature of the 3.5-keV line (and another at 4.5 keV) in NuSTAR remain unknown, \citet{perez} set deliberately conservative limits on the line fluxes that could be due to new signals.

In a recent paper, \citet{nero} reported a 11 $\sigma$ detection of the 3.5 keV line in NuSTAR observations of the CCLS field and the CDFS. They observed the same areas of sky observed here, for 
a comparable exposure time, taking advantage of the fact that the NuSTAR detector which was not shielded from indirect light, was able to effectively survey a total sky area of 37.2 deg${^2}$ viewed by a 
13$\arcmin\times$13$\arcmin$ detector area. This obviously provides increased leverage compared to telescopes sensitive to focused photons only. Interestingly, the line has been detected by \citet{wik} but 
no hypothesis has been put forward for its origin. In fact, the line has been flagged as instrumental. Chandra and NuSTAR have the same collecting area at 3.5 keV and the exposures used in these two papers are comparable. We can therefore, directly compare the two results by transforming the observed fluxes into surface brightness (S) under the assumption that the line flux is homogenous over the 37.5 deg$^2$, however for NuSTAR (S$_{3.5,Nu}$) we have to take into effect the boosting factor introduced by the non-focused component of the signal so that:
\begin{equation}
\label{eq:nu}
S_{3.5,Nu}=F_{3.5,Nu}/(\kappa(E)* 1.43\times10^{-5}),
\end{equation}
\noindent
where F$_{3.5,Nu}$ is the  flux of the line observed by NuSTAR and $\kappa(E)$ is the energy dependent $boosting factor$ for the 
NuSTAR measured diffuse indirect background. This takes into account the fact that the effective surveyed area is much larger
than the area sensitive to focused photons. 

At 3.5 keV, \citet{nero} report $\kappa(E)\sim$7.5 and the field of view (f.o.v.) of 
the Cd Zn Te detector is 1.43$\times10^{-5}$ sr while  ACIS-I's f.o.v. is   2.42$\times10^{-5}$ sr.  Considering this we find 
S$_{3.5,Nu}$=0.093$\pm{0.023}$ ph/cm$^{2}$/s/sr and  S$_{3.5,Nu}$=0.069$\pm{0.012}$ ph/cm$^{2}$/s/sr  with data taken in the shadow 
of the earth and illuminated by the Sun, respectively and S$_{3.5,Ch}$=0.042$\pm{0.017}$ ph/cm$^{2}$/s/sr with Chandra.

Our measurements are thus  marginally consistent  with NuSTAR's by \citet{nero}
thus, it  is possible  that Chandra and NuSTAR are observing the same cosmic source of 3.5 keV photons. 
However if the flux of the line is as measured by NuSTAR, we would have detected the line at least 5$\sigma$.
However, it is worth noting that the calibration of the effective area of NuSTAR in that energy band 
is very unstable (as per information from the NuSTAR Calibration team) and a 2\% spike could be introduced by the 
fact that during the calibration the control points for the Crab fitting are at 3.3 and 3.68 keV, the Crab 
and hence the response has been corrected between these two energies with a straight line.

If the line is not an artifact, the NuSTAR detection is  $\sim$3 times more significant because they collected 10 times more photons than Chandra did.  Assuming a consistency between the measurements (even if marginal), 
given the differences in satellite orbits and detectors, means an instrumental or cosmic ray origin for the signal is unlikely. The intensity of line is the same both with the spacecraft illuminated by the Sun and in the shade of the Earth.
Moreover, $Chandra$ observations were taken 
over $\sim$ 15 years while NuSTAR data were obtained in just the past 3 years which argues against such transient causes such as the solar wind.  However the energy of the 
line is remarkably consistent with the two observations, taken with two different instrumental setups\footnote{ACIS-I is a silicon CCD while the imagers of NuSTAR are two 
Cadmium-Zinc-Telluride detectors}, under different geomagnetic conditions and at completely different times, suggests an extrinsic source for the detected line. 
\citet{hitomi16} speculated that the line might be a feature of CCD detectors but this would not account for the NuSTAR detection with CdZnTe detectors. 

 Moreover, a recent analysis of the $Chandra$ PIB by \citet{barta} did not find any residuals nor emission lines 
 between 3 and 5.8 keV. While we cannot exclude further unaccounted and as yet unknown effects introduced by the mirrors or the CCD, based on this concordance  the instrumental origin seems to be less likely with multiple detections in the data taken with different instruments and under 
 different conditions. A further source of concern is the  contamination of  ACIS optical blocking filter by a deposit of hydrocarbons. This effect has been known for 
 many years and well understood. Moreover, while this effect is dramatic in the soft bands, it is small above 3 keV and we consider it negligible. 
 We also investigated the possibility that Tin whiskers (crystalline structures of tin growing when tin coatings are used as a finish) might be implicated, since Sn$_{50}$ presents 
 energetic transitions in L shells  around 3.5 keV. However, consulting with the Chandra engineering team suggests that the amount of tin is relatively small but we couldn't estimate its contribution to our observations. Still, further 
 calibrations, and deeper studies of the spectral dependence of the instrument response are needed and  will be important for firmly establishing the reality (or not) of this emission 
 feature. In particular, we would recommend deeper integrations of the stowed background.\\ 
With this analysis we can affirm that unless the Chandra effective area calibration has  problems at 3.5 keV that remain undetected despite substantial attention to this energy, we can exclude an instrumental origin for the line.  We now proceed to discuss possible physical mechanisms that can  produce an emission line at 3.5 keV. 
 
\subsection{The Iron Line Background}

 Regardless of the nature of the search, we know that when observing the CXB, we are witnessing the accretion history onto Super Massive Black Holes across cosmic time. There is evidence that a large fraction of the   accretion in the universe occurs in an obscured phase \citep[see e.g.,][]{gilli07,treist05}. One characteristic feature of such a phase of accretion is a strong Fe K$_\alpha$ 6.4 keV emission line.  Such an emission line has been significantly detected in stacked spectra of AGN  divided into redshift bins \citep[see e.g.][]{brusa05,falocco,pooja}, with a very intense contribution from sources  at  z$\sim$0.7-.0.9 (i.e.  Fe K$_\alpha$ redshifted to 3.5 keV) where the cosmic AGN activity was near its peak. However, the  CXB spectrum contains the emission from AGN from all redshifts and its intensity is modulated by the redshift distribution of the sources and their luminosity distance. \citet{gilli98,gilli99} modeled this emission and found that the the redshift  distribution smooths this signal into an 'inverse edge'-shaped feature between 2 and 4 keV. The intensity of such a feature is a few percent above the continuum at about 3.5 keV, however since the redshift distribution of the resolved sources is not smooth but it shows spikes due to the presence of large scale structure, the feature appears near or at the energy of such spikes. Both COSMOS and the CDFS do not show prominent spikes in their AGN redshift distribution around z$\sim$0.8 \citep{luo,marchesi} this, together with the lack an 'inverse edge' feature in the spectrum, we safely state that this  scenario is unlikely an can be excluded.

\subsection{3.5 keV  line from S XVI Charge exchange}
Gu et al. (2015) suggested that the 3.5 keV line could be attributed to Charge Exchange (CX) between neutral Hydrogen and bare Sulfur ions. This collision leads to the full Lyman series of transitions in S XVI, with a strong Ly$_{\alpha}$ at 2.62 keV and, crucially, enhanced high $n$ transitions around Ly$_{\eta}$ and Ly$_{\theta}$ (i.e. $n=8\rightarrow1, 9\rightarrow1$) transition. These enhanced high $n$ lines are the indicator of CX, driven by capture into the high $n$ shells which does not occur during electron impact collisional excitation. Significantly for this work, these lines lie in the 3.4--3.45keV energy band.  
The exact ratios of the lines in the Lyman series depends on the exact $n$ and $l$ shell into which the electron is captured. In particular, the $l$ shell is very sensitive to the collision energy, although calculations of the relative cross section are sparse and highly likely to disagree. We have used data from the AtomDB Charge Exchange (ACX) model \citep{smith} to obtain the line energies and relative intensities shown in Table 4. In this case we have used ACX model 8, which is the separable $l$ distribution and the weighted $n$ distribution (described in Smith et al. 2012). This corresponds to relatively low center of mass velocity ($\lesssim$ 1000km/s) which is appropriate for a thermal plasma such as this one, however the results do not change significantly if other distributions are used instead.

In all of these observed scenarios, the intensity of the Ly$\alpha$ line is 5 times that of the 3.4-3.45keV line complex. We do not detect a line with an energy consistent with 2.62 keV, although we can determine an upper limit for its intensity at $<2.98\times10^{-6}$ ph/cm$^2$. By assuming that all of the $\sim$3.5 keV emission is produced by S XVII CX, and considering the energy resolution of Chandra (of the order 150 eV) and NuSTAR (400 eV), we test the hypothesis of Gu et al. (2015) and Shah et al. (2016) that we are seeing a blend of all the possible transitions around 3.4-3.45 keV. Although, the energy of the line detected here is clearly in tension with the predictions for S XVII CX, the discrepancy just might be a consequence of the energy resolution of the instrument. 

From the values in Table 4, we expect a line ratio I$_{3.45}$ /I$_{2.62}$ of $\le$ 0.2, where I$_{3.45}$ is the intensity of the 3.45 keV line system. In our case the ratio is $>$0.34 which rules out CX together with a discrepant energy. In addition, any signal at 2.62 keV, that we can interpret here as the n=2$\rightarrow$1 S XVII transition can also be attributed to the daughter lines of the instrumental feature at 2.1 keV. Any such contribution would, in effect, raise the observed ratio, making CX  less likely. In addition, the CX process should also produce a significant Ly$_{\beta}$ line at 3.106keV: we do not observe no such line, but we can only place an upper limit (see Table \ref{tab:sulfur}).
Another possible CX transition that occurs near 3.5 keV is the Ar XVIII n=2$\rightarrow$1 transition at 3.32 keV, where we do not detect any line nor do we see any evidence of higher n shell transitions from this ion. According to these measurements and atomic calculations, 
we can rule out  that the totality of the 3.5 keV line  flux measured here is  produced by CX.

\begin{table}
\begin{center}
\footnotesize
\caption{Predicted S XVI Charge exchange transitions lines. \label{tab:sulfur} }

\renewcommand{\arraystretch}{1.5}
\begin{tabular}{lcccc}
\hline\hline
	 Transition & Energy &  I$_{E}$/I$_{2.62}$ & I(E)\\
	&                      keV   & 10$^{-6}$ ph/cm$^2$.  \\
	\hline\hline

 2$\rightarrow$1 & 2.621 &1.0 &$<$2.98\\
3$\rightarrow$1 &3.106 &0.142 & $<$1.45 \\
4$\rightarrow$1 &3.276& 0.050 &$<$0.64 \\
5$\rightarrow$1 &3.354& 0.025 &$<$0.51\\
6$\rightarrow$1 & 3.397 &0.016 &$<$0.64\\
7$\rightarrow$1 &3.423 &0.011&1.02$^a$ \\
8$\rightarrow$1& 3.434& 0.120& 1.02$^a$\\
9$\rightarrow$1 &3.451 &0.074&1.02$^a$ \\
   \hline
\end{tabular}
\tablenote{ assuming the detected 3.5 keV flux}.
\end{center}
\end{table}

\subsection{3.5 keV line from dark matter decay}

One of the possible interpretations of the detection of the 3.5 keV emission line is the decay of sterile neutrinos into a neutrino and a X-ray photon  \citep{pal}.
If the emission originates from DM decay, then the line flux would be proportional to the amount of matter along the line of sight over the field-of-view. In the present case, we would expect the Milky Way dark matter halo to dominate the local signal \citep{rr06}. With this data set, we sample the DM halo distribution along the line of sight and therefore, the emission seen should scale with amount of mass sampled.

 \citet{bo14}, detected the 3.5 keV line in the direction of the GC. The observed fields presented here 
lie at an aperture angle $\theta$ with respect to the GC. If our detected signal comes from DM decay within the MW halo then its intensity should be:
\begin{equation}
\label{eq:profile}
I_{DM}(\theta)=I_{DM,GC}\frac{\int \rho_{DM}[r(l,0^{\circ})]dl\,d\Omega}{\int \rho_{DM}[r(l,\theta)]dl\,d\Omega}         \\  
\end{equation}
where, $I_{DM}(\theta)$ is the DM decay signal at aperture angle $\theta$ from the GC;
$I_{DM,GC}$ is the DM decay signal from the GC  ($\theta$=0);
  $\rho$(r) is the DM density profile; $l$ is the distance along the line of sight; $r$ and $\theta$ are the physical and angular 
distance from the center of the galaxy, respectively. The three quantities are related via 
\begin{equation}
r(l,\theta)= \sqrt{l^2+d^2-2 ldcos(\theta)}
\end{equation}
where $d$ is the distance of the earth from the GC.  We note that the distance and MW DM profile parameters and shape
are still highly debated \citep{bh}. 

Assuming that all the intervening dark matter is associated with a cold component that can be modeled with an NFW profile \citep{navarro1997} given by: 
\begin{equation}
\rho_{DM}=\frac{\rho^{*}}{x(1+x)^2}
\end{equation}
where $x=r/r_H$; here we adopt the parameters measured by Nesti \& Sallucci (2013):
and therefore use $d$=8.02$\pm{0.2}$ kpc, $r_H$=16.1$^{+12.2}_{-5.6}$, 
$\rho^{*}$=13.8$^{+20.7}_{-6.6}\times10^6$ M$_\odot$/kpc$^{3}$ and $I_{DM,GC}$=0.63$\pm{0.11}$  ph/s/cm$^2$/sr.
Using Eq. \ref{eq:profile} we calculated, with Monte Carlo integration, the 1$\sigma$ and 2$\sigma$ confidence levels of the 
flux from DM decay along the line of sight as a function of the angular distance from the GC. 
This is shown in Figure \ref{fig:nfw}, wherein we overplot our measurement and the NuSTAR measurement.
The two fields investigated here are basically at the same angular distance from the GC of $\theta\sim$115 deg. Remarkably, our 
measurements are consistent at the 1$\sigma$ level with such a profile.  This means 
the ratio of fluxes at $\theta$=115 and $\theta$=0 is consistent with the NFW DM decay model. We also point out that we
assumed that Sgr A$^{*}$ coincides with the centre of the MW DM halo.

\begin{figure}
\begin{center}
\includegraphics[width=1\columnwidth]{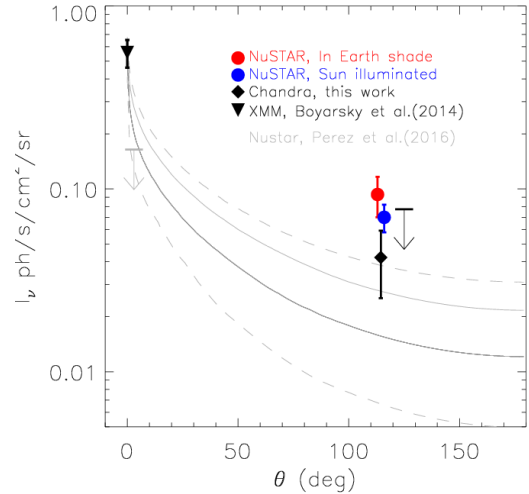} 
\caption{1$\sigma$ ($continuous~line$) and 2$\sigma$ ($dashed~line$) limits on the  expected 3.5 keV line flux as function of the angular distance from the GC by assuming a NFW profile with 
parameters from \citet{nesti} and DM flux at $\theta$=0 from \citep{bo14}. The profile is compared with our measurements from the deep fields ($black~filled~circles$) and with the NuSTAR results  
($red/blue~filled~circles$). The $downward-black$ arrow represents the 3$\sigma$ limit derived from simulations. \label{fig:nfw} 
}
\end{center}
\end{figure}

In terms of constraints on the number of neutrino species (allowing one additional species of a 
sterile neutrino along with the 3 other usual flavors), \citet{plank} report that with the CMB temperature data alone 
it is difficult to constrain $N_{\rm eff}$, and data from Planck alone do not rule out $N_{\rm eff} = 4$. 
At the 95\% C.L. combining Planck + WMAP + high l experiments they obtain $N_{\rm eff} = 3.36^{+0.68}_{-0.64}$. The Planck collaboration 
has only investigated an eV mass sterile neutrino as a potential additional species. So other than saying that $N_{\rm eff} = 4$ is 
permitted, there are no concrete CMB constraints on keV sterile neutrinos.

Performing the line integral through the halo of the Milky Way taking into account the f.o.v and given that all 3 deep fields
included in this analysis are at roughly 115 degrees, we compute the surface mass density along the line of sight. 
Similar to our assumption adopted above, the MW halo is once again modeled with an NFW profile and the current best-fit parameters 
are adopted from \citet{nesti}. Using the formulation developed in \citet{aba07}, we use the measured flux in the line to constrain the mixing 
angle $\sin^2{ 2\theta}$. Although we use the integrated surface mass density of dark matter in the Milky Way halo integrated out
to the virial radius, the dominant contribution comes from the inner region - from within a few scale radii - of the density profile 
due to the shape of the NFW profile. Using the higher bound and the lower bound estimates for the total mass of the Milky Way,
we obtain the following values for $\Sigma$ the integrated surface mass density of DM:
\begin{eqnarray}
\begin{split}
\Sigma_{\rm DM, High} = 0.0362\,{\rm gm\,cm^{-2}}; \\
\Sigma_{\rm DM, Low} = 0.0109\,{\rm gm\,cm^{-2}}.\\
\,
\end{split}
\end{eqnarray}

\begin{figure*}
\begin{center}
\hbox{\includegraphics[width=1\columnwidth]{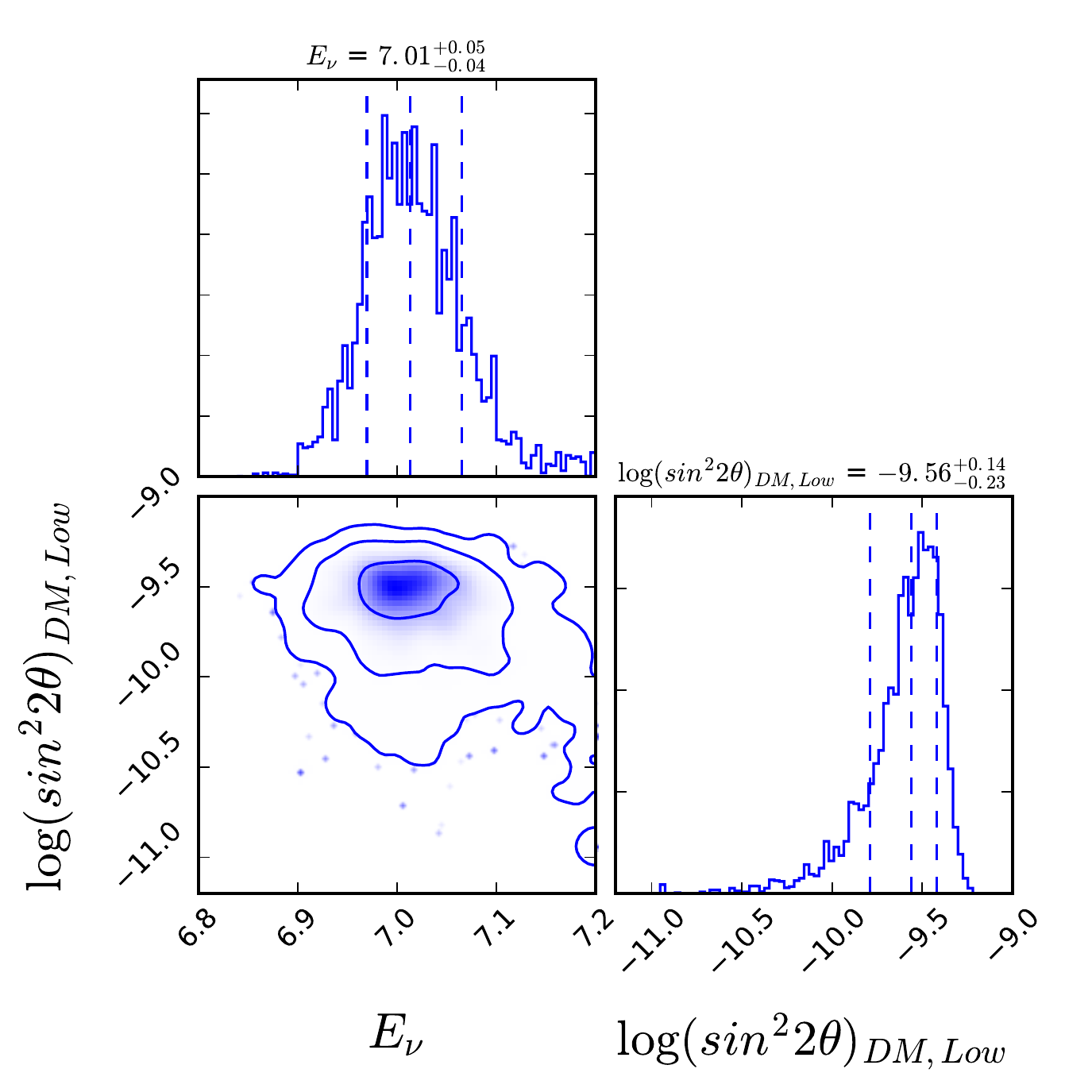}
 \includegraphics[width=1\columnwidth]{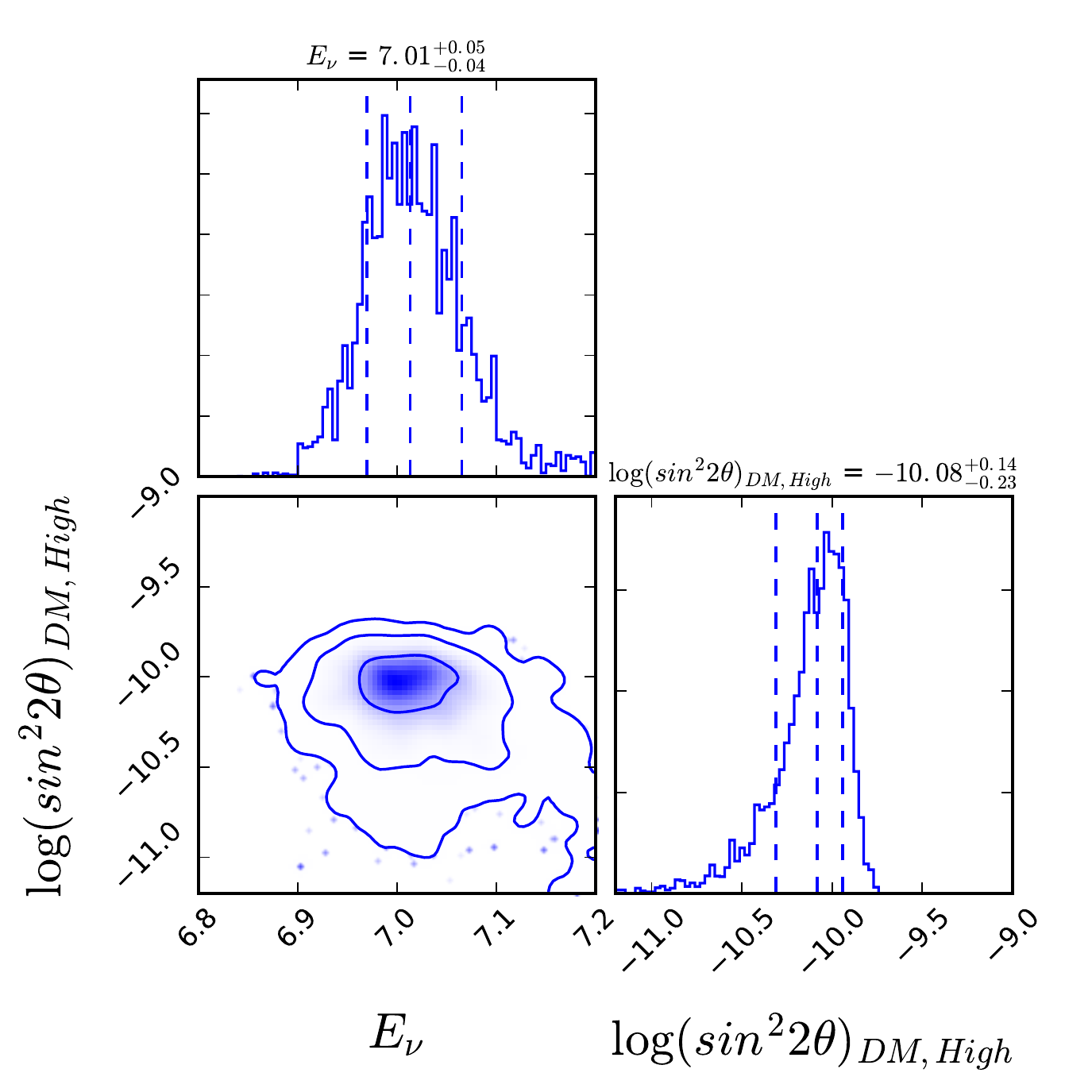}

}
\caption{\label{fig:SN} The confidence contours derived for sterile neutrino parameters E$_{\nu}$ and $\sin^2{2 \theta}$  from our MCMC
for the cases of low  and high  integrated surface mass density of DM ($\Sigma_{DM}$) $left$ and $right$, respectively.}
\end{center}
\end{figure*}

Using these values and the equation:
\begin{eqnarray}
\begin{split}
\sin^2{2 \theta} \times (\frac{m_{\nu}}{1\,{\rm 1\,keV}})^4 \times \frac{\Sigma_{\rm DM}}{\rm gm\,cm^{-2}} =\\
 (\frac{I_{\nu}}{1.45 \times 10^{-4}})\,{\rm photons\,cm^{-2}\,s^{-1}\,{arcsec^{-2}}},\\
 \,
\end{split}
\end{eqnarray}

we obtain that $\sin^2{ 2\theta}_{\rm DM, High} = 0.83^{+0.34}_{-0.31}  \times 10^{-10}$ and $\sin^2{ 2\theta}_{\rm DM, Low} = 2.75^{+1.13}_{-1.04} \times 10^{-10}$. 
The confidence contours for the sterile neutrino parameters new summarized in Figure \ref{fig:SN}.
Furthermore, we can now estimate the lifetime $\tau$ for this sterile neutrino species, using equation 2 of Boyarsky et al. (2015):
\begin{eqnarray}
\tau_{\rm DM} = 7.2 \times 10^{29}\,{\rm sec} (\frac{10^{-8}}{\sin^2{2 \theta}})\,(\frac{1\,{\rm keV}}{m_{\nu}})^5
\end{eqnarray}
and find that it is $\tau_{\rm DM, High} = 5.16^{+3.56}_{-1.42} \times 10^{27}$ sec and $\tau_{\rm DM, Low} = 1.55^{+1.06}_{-0.43} \times 10^{27}$ sec respectively.
These mixing angle estimates are in very good agreement with Figures~13 and 14 of Bul14.  They can also be overplotted
and seen clearly to be consistent with Figure~3 of \citet{iakubovskyi15}.

However, despite concordance with parameters extracted from other observational constraints obtained from
X-ray data of stacked galaxy clusters and the Galactic center, due to the significance of our detection only at  $\sim$3$\sigma$ level, we cannot
conclusively claim that this observed 3.51 keV line originates from decaying dark matter.  
It  would require a non-detection with at least 100 Ms of $Chandra$ observations to rule out this hypothesis.

\section{Summary}

 In this paper  we perform a systematic  search for an  emission feature at $\sim$3.5 keV in the spectrum of the CXB with extremely deep {\em Chandra} integration time. We find evidence of a feature with a  significance  of 2.5-3$\sigma$, depending  on the  statistical treatment of the data, respectively. 
The evaluation of the significance of the line is further complicated by the complexity of the model and the weak nature of the signal. In particular, to estimate  the relation between $\Delta\chi^2$  and P is complicate because of model misspecification. Additionally, regardless of the significance of the feature we are able to place
a 3$\sigma$ upper-limit to the line intensity.  
Examining the sources of possible origin for this feature, we conclude that the line does not have a clear known instrumental origin.
 The intensity and the energy of the line is consistent with earlier measurements that were interpreted as decay of $\sim$7 keV sterile neutrino and the decay rate found here is in remarkable agreement with previous work. We can interpret the signal  as DM decay along the line of sight  in the Milky Way halo. 

We also investigate  the scenario wherein the 3.5 keV flux is produced by charge exchange between neutral Hydrogen with bare Sulfur ions.
We conclude that  all the 3.5 keV flux cannot be produced by charge excange.
We also discuss a scenario, in which the line could be produced by a blend of redshifted iron lines from AGN by large scale structures that spike at z$\sim$0.8. This interpretation would be consistent with predictions for the iron line background but not a) with cluster measurements (Bul14) and b) with the lack of prominent spikes in the redshift distribution at that redshift.
We can  conclude that charge exchange and the Iron Line background together cannot produce more that 1.85$\times$10$^{-6}$ ph/cm$^2$/s at 3.5 keV.
So far, the 3.5 keV  line is the only feature detected from 4 independent instruments that is interpretable as DM decay ($Chandra$, XMM-$Newton, ~Suzaku$ and $NuSTAR$)  with more than one  $>$5 $\sigma$ detection in a variety of DM dominated objects. Given the amount of data available in the archives, an intensive data mining exercise of X-ray spectra is an extremely cost- and time-effective method to rule out or confirm the contribution of sterile neutrinos to DM. The nature of dark matter is a key unsolved problem in cosmology and at the moment we seem to be at an impasse in terms of both direct and indirect detection experiments \citep[see e.g.][]{ack,2016arXiv161205949I}. Therefore, further even more careful analysis of existing X-ray observations is warranted and crucial. In the future, X-ray calorimeters on board of {\it XARM} (X-ray Astronomy Recovery Mission),  {\it Athena} or  the Micro-X sounding rocket \citep{figue} will  greatly improve our understanding of the origin of the 3.5 keV feature  given their capability for high precision spectroscopy.
%

\section{Acknowledgements}
NC acknowledges the Yale University YCAA Prize  fellowship 
postdoctoral program. PN acknowledges a Theoretical and Computational Astrophysics Network grant with award number 1332858 from the National Science Foundation and 
thanks the Aspen Center for Physics, which is supported by the National Science Foundation grant PHY-1066293, where
this work was done in part.
EB acknowledges support from NASA grants NNX13AE77G

\end{document}